\documentclass[%
reprint,
superscriptaddress,
showpacs,
 amsmath,amssymb,
aps,
]{revtex4-1}
\usepackage{mathrsfs}
\usepackage{color}
\usepackage{epsfig}
\usepackage{graphicx}
\usepackage{dcolumn}
\usepackage{bm}
\usepackage[mathlines]{lineno}

\begin{document}
\title{Converting normal insulators into topological insulators via tuning orbital levels}
\author{Wu-Jun Shi}
\affiliation{State Key Laboratory of Low-Dimensional Quantum Physics, and Department of Physics, Tsinghua University, Beijing 100084, People's Republic of China}
\affiliation{National Laboratory of Solid State Microstructures, and Department of Physics, Nanjing University, Nanjing 210093, People's Republic of China}
\author{Junwei Liu}
\affiliation{Department of Physics, Massachusetts Institute of Technology, Cambridge, MA 02139, USA}
\author{Yong Xu}
\email{yongxu@mail.tsinghua.edu.cn}
\affiliation{State Key Laboratory of Low-Dimensional Quantum Physics, and Department of Physics, Tsinghua University, Beijing 100084, People's Republic of China}
\affiliation{Collaborative Innovation Center of Quantum Matter, Tsinghua University, Beijing 100084, People's Republic of China}
\affiliation{Department of Physics, McCullough Building, Stanford University, Stanford, California 94305-4045, USA.}
\author{Shi-Jie Xiong}
\affiliation{National Laboratory of Solid State Microstructures, and Department of Physics, Nanjing University, Nanjing 210093, People's Republic of China}
\author{Jian Wu}
\affiliation{State Key Laboratory of Low-Dimensional Quantum Physics, and Department of Physics, Tsinghua University, Beijing 100084, People's Republic of China}
\affiliation{Collaborative Innovation Center of Quantum Matter, Tsinghua University, Beijing 100084, People's Republic of China}
\author{Wenhui Duan}
\email{dwh@phys.tsinghua.edu.cn}
\affiliation{State Key Laboratory of Low-Dimensional Quantum Physics, and Department of Physics, Tsinghua University, Beijing 100084, People's Republic of China}
\affiliation{Collaborative Innovation Center of Quantum Matter, Tsinghua University, Beijing 100084, People's Republic of China}

\date{\today}
\begin{abstract}
Tuning the spin-orbit coupling strength via foreign element doping and/or modifying bonding strength via strain engineering are the major routes to convert normal insulators to topological insulators. We here propose an alternative strategy to realize topological phase transition by tuning the orbital level. Following this strategy, our first-principles calculations demonstrate that a topological phase transition in some cubic perovskite-type compounds CsGeBr$_3$ and CsSnBr$_3$ could be facilitated by carbon substitutional doping. Such unique topological phase transition predominantly results from the lower orbital energy of the carbon dopant, which can pull down the conduction bands and even induce band inversion. Beyond conventional approaches, our finding of tuning the orbital level may greatly expand the range of topologically nontrivial materials.
\end{abstract}

\maketitle

Topological insulators (TIs), as new quantum states of materials characterized by insulating bulk and metallic surface states that are topologically protected against backscattering, have attracted enormous interest in recent years due to their novel electronic properties\cite{Hasan_Kane_Rev, Qi_Zhang_Rev}. So far many TI materials have been theoretically predicted and/or experimentally identified, including HgTe quantum wells\cite{2DTI1, 2DTI2}, Bi$_2$Se$_3$-class TIs\cite{Bi2Se3_cal_1, Bi2Se3_exp_3, Bi2Se3-exp-com}, TlBiSe$_2$ TIs\cite{TlBiSe-com}, half-Heusler TIs\cite{heusler-com}, \emph{etc}. All of these materials inherently have an inverted band structure and thus are topologically nontrivial. However, there exist a much wider range of other materials which intrinsically are topologically trivial but potentially can be changed into TIs. Such large class of materials, if applicable for TI-related research and applications, would greatly facilitate the development of condensed matter physics and materials science.

How to convert conventional materials (or more specifically, semiconductors) into TIs is a crucial problem yet to be solved. A few approaches have been proposed and developed for the purpose. For example, topological phase transitions in normal insulators can be induced by manipulating the crystal lattice via external strain\cite{NM-all, halide1, strain1, strain2, strain3, strain-add} and chemical doping/functionalization\cite{TlBiSSe1, TlBiSSe1_2,BiInSe,strain3_xy, strain4_sc, strain-add3, SOC-reduce}, or by tuning the electronic structure via electric field\cite{elec-filed} and quantum confinement\cite{quan-confine}. The general physical picture behind these approaches is either to tune the spin-orbit coupling (SOC) strength or to change the bonding strength of the normal insulators, so as to induce the band inversion required.

In this work, we reveal an alternative, physically-distinct approach to achieve TI states from conventional materials, where doping heavy elements with large SOC strength and applying large strain are not requisite any more. Based on the tight binding Hamiltonian and first-principles calculations, we demonstrate that topological phase transition of normal insulators can be successfully realized by tuning the orbital levels (rather than SOC strength) via doping. This topological phase transition is originated from the difference in orbital levels between the host and dopant atoms. As examples, we show that the perovskite-type compound CsSnBr$_3$ can be converted from normal insulators to three-dimensional (3D) strong TIs by substituting Ge or Sn with C. The lower $p$ orbital level of C ($2p$) than that of Ge ($4p$) or Sn ($5p$) pulls down the conduction bands and could lead to an interchange between the original conduction band minimum (CBM) and valance band maximum (VBM), i.e., these normal insulators are converted to TIs.

\begin{figure}[tpb]
\includegraphics[width=0.35\textwidth]{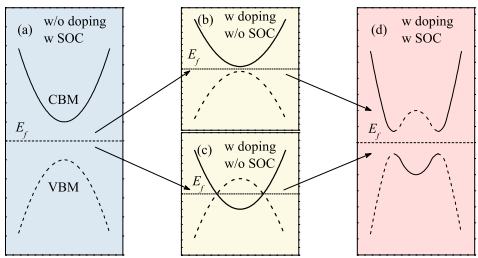}
\caption{\label{fig:schem} (Color online) Schematic illustration of the doping and SOC induced band inversion. (a) The original bands are not inverted. (b)/(c) The CBM and VBM shift toward each other induced by doping is less/larger than the band gap in (a). Note that a band inversion occurs in (c). (d) The final bands are inverted. For the path (a)-(b)-(d), the doping reduces the band gap, and the SOC induces the band inversion, while for path (a)-(c)-(d), the doping induces the band inversion, and the SOC opens a band gap. The solid and dashed curves denote the subbands that have opposite parities. The Fermi level ($E_f$) is denoted by dashed line. }
\end{figure}

We first consider a generic tight-binding Hamiltonian:
\begin{equation*}
\mathcal{H} = \sum_{\mu\vec{R}}\epsilon_{\mu} c_{\vec{R}}^{\mu \dag} c_{\vec{R}}^{\mu} + \sum_{\mu \nu\vec{R} \vec{R^\prime}} t_{\vec{R} \vec{R^\prime}}^{\mu \nu} c_{\vec{R}}^{\mu \dag}c_{\vec{R^\prime}}^{\nu} + \sum_{\vec{R}}\lambda_{\vec{R}}\vec{L}_{\vec{R}}\cdot\vec{s}_{\vec{R}},
\end{equation*}
where $\epsilon_{\mu}$ is the on-site energy of orbital $\mu$, $c_{\vec{R}}^{\mu \dag}$ ($c_{\vec{R}}^{\mu}$) is the fermion creation (annihilation) operator of orbital $\mu$ at site $\vec{R}$; $t_{\vec{R} \vec{R'}}^{\mu \nu}$ is the hopping integral between orbital $\mu$ at site $\vec{R}$ and orbital $\nu$ at site $\vec{R^\prime}$; $\lambda_{\vec{R}}$ is the SOC strength. From the tight-binding view, the electronic band structures are determined by the on-site energy $\epsilon$, the hopping integral $t$ and the SOC strength $\lambda$. Therefore, the topological phase transition may be realized by tuning these three parameters of certain normal insulators. Actually it has been extensively shown that doping heavy atoms can be used to increase the SOC strength\cite{TlBiSSe1, TlBiSSe1_2,BiInSe} or doping lighter atoms to decrease the SOC strength, as demonstrated in the example of doping Bi with Sb \cite{SOC-reduce}, while applying the external strain can change the interatomic distances which can affect the atomic wavefunction overlap and also the hopping integrals\cite{halide1, strain2, strain3, strain3_xy, strain4_sc}. In contrast, herein, we focus on the role of the on-site energy change (by chemical doping) in the topological phase transition. Fig.~\ref{fig:schem} schematically shows the relevant topological phase transition mechanism: due to the difference in orbital level (i.e., the on-site energy) between the substitutional dopant and host atoms, the CBM and VBM shift towards each other owe to doping, causing a reduction even disappearance of the band gap; SOC then further induces a band inversion or band gap reopening.

\begin{table}[tpb]
\caption{\label{tab:table}The PBE calculated equilibrium lattice constant ($a_0$), global band gap without SOC ($E_{\tt g}$), with SOC ($E^{\tt soc}_{\tt g}$), and band gap ($E^{\tt soc}_{\tt TRIM}$) at $R$ (pure) or $\Gamma$ (doped) point of pure CsGeBr$_3$, pure CsSnBr$_3$, CsGe$_{0.875}$C$_{0.125}$Br$_3$, and CsSn$_{0.875}$C$_{0.125}$Br$_3$. ``$-$'' presents the inverted band gap.}
\begin{ruledtabular}
\begin{tabular}{ccccc}
                                & $a_0$ (\AA) & $E_{\tt g}$ (meV) & $E^{\tt soc}_{\tt g}$ (meV) & $E^{\tt soc}_{\tt TRIM}$ (meV)\\
\colrule
CsGeBr$_3$                      & 5.604       & 745.2             & 583.4                       & 583.4   \\
CsGe$_{0.875}$C$_{0.125}$Br$_3$ & 11.085      & 0.8               & $-$16.1                     & $-$31.6 \\
CsSnBr$_3$                      & 5.883       & 625.9             & 279.2                       & 279.2   \\
CsSn$_{0.875}$C$_{0.125}$Br$_3$ & 11.578      & 0.0               & $-$16.1                     & $-$35.3 \\
\end{tabular}
\end{ruledtabular}
\end{table}

In order to prove the above strategy, we perform the first-principles calculations. As well known, in the $ABX_3$-type perovskite structures, the electronic states near the Fermi level are mainly contributed by the $B$-site and $X$-site atoms\cite{halide1,mine1}. Therefore, these perovskite structures with the $B$-site or $X$-site substitution by foreign atoms with different orbital levels are suitable to tune the bands near the Fermi level\cite{mine1}. Herein, the halide perovskite-type compounds CsGeBr$_3$, and CsSnBr$_3$, which might be converted into topological states under external strain\cite{NM-all, halide1} are selected to demonstrate the concept.

It is worthwhile to note that halide perovskite-type compounds can exhibit rich structural phases under different conditions of pressure and temperature. The low temperature phase of CsGeBr$_3$ is rhombohedral\cite{CGB-temperature-pressure} and that of CsSnBr$_3$ is tetragonal, or monoclinic \cite{CSB-all-phase,CSB-temperature-BG}. All these materials will transit into the cubic phase when increasing temperature. The transition temperature $T_{\tt C}$ is $\sim$~510~K for CsGeBr$_3$ \cite{CGB-temperature-pressure}, and is around room temperature ($\sim$292~K) for CsSnBr$_3$\cite{CSB-all-phase, CSB-temperature-BG}. In fact, a lower $T_{\tt C}$ could be achieved by applying an external pressure. For instance, the cubic phase becomes stable at room temperature under an external pressure of $\sim$1~GPa\cite{CGB-temperature-pressure}. On the other hand, a structural phase that is originally metastable in its freestanding form could be stabilized when grown on the substrate, as shown by a recent experiment of SnSe on Bi$_2$Se$_3$ substrate\cite{mbe}. Thus, for the sake of simplicity and without loss of generality, we only consider the cubic phase of CsGeBr$_3$ and CsSnBr$_3$ in the following.

The first-principles calculations are performed using the Vienna {\it ab initio} simulation package (VASP)\cite{vasp1} with the plane-wave basis. The interactions between the valence electrons and ion cores are described by the projector augmented wave (PAW) method\cite{paw}, and exchange-correlation potential is formulated by the generalized gradient approximation (GGA) with the Perdew-Burke-Ernzerhof (PBE) scheme\cite{pbe}. The Heyd-Scuseria-Ernzerhof (HSE)\cite{hse} hybrid functional is employed to check the results. The $\Gamma$-centered $k$ points are used for the first Brillouin zone sampling. The plane-wave basis cutoff energy is set to 500 eV. The structures are optimized until the forces on atoms are less than 5~meV/\AA. For pure CsGeBr$_3$ and CsSnBr$_3$, a dense $10\times10\times10$ grid of $k$ points is used. To simulate doping effect, a $2\times2\times2$ supercell is employed with one Ge or Sn substituted by one C (CsGe$_{0.875}$C$_{0.125}$Br$_3$ or CsSn$_{0.875}$C$_{0.125}$Br$_3$); and a $5\times5\times5$ grid of $k$ points is used. The fully relaxed lattice constants of pure and doped structures are listed in Table~\ref{tab:table}, and will be used in subsequent calculations. Note that the lattice constants of pure structures are consistent with the values in Ref.~\cite{NM-all}.

\begin{figure}[tpb]
\includegraphics[width=0.35\textwidth]{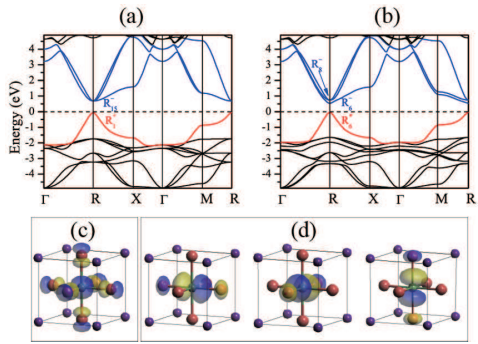}
\caption{\label{fig:CsGeBr$_3$_111_bs_dos} (Color online) The calculated band structures of pure CsGeBr$_3$ without (a) and with SOC (b); the real part of the wave functions of the $R_1^+$ (c) and $R_{15}^-$ (d) states without SOC, the blue (yellow) isosurfaces represents ``$+$'' (``$-$'') sign of the wavefunctions. The Fermi level is set to zero and indicated by dashed line.}
\end{figure}

Firstly, we analyse the electronic structure and topological properties of CsGeBr$_3$ and CsSnBr$_3$. As an example, the calculated band structures of CsGeBr$_3$ are shown in Fig.~\ref{fig:CsGeBr$_3$_111_bs_dos}. It can be seen that without SOC, CsGeBr$_3$ has a direct band gap at the $R$ point [Fig.~\ref{fig:CsGeBr$_3$_111_bs_dos}(a)]. A further analysis of the wave functions and their symmetries reveals that the VBM states consist of Br $4p$ and Ge $4s$ states and possess the R$_1^{+}$ symmetry with the even parity [Fig.~\ref{fig:CsGeBr$_3$_111_bs_dos}~(c)], while the CBM states mainly come from Ge $4p$ and have $R_{15}^{-}$ symmetry with the odd parity [Fig.~\ref{fig:CsGeBr$_3$_111_bs_dos}~(d)]. By taking SOC into account [Fig.~\ref{fig:CsGeBr$_3$_111_bs_dos}(b)], GsGeBr$_3$ is still a narrow-gap semiconductor with direct gap at $R$ point, while the $R_{15}^{-}$ states (the CBM) split into two-fold degenerate $R^{-}_6$ states and four-fold degenerate $R^{-}_8$ states, and the VBM has the two-fold $R_{6}^{+}$ symmetry. Meanwhile, the band gap decreases from 745.2 to 583.4~meV (Table~\ref{tab:table}). The electronic structures of CsSnBr$_3$, CsSnCl$_3$ and CsSnI$_3$ are similar to that of CsGeBr$_3$, except for the magnitude of band gap (see Table~\ref{tab:table}). By calculating the parities of all the occupied states at the time-reversal-invariant momentum (TRIM) points (i.e., $\Gamma$, $R$, $X$, and $M$ points), we find that pristine CsGeBr$_3$, and CsSnBr$_3$ are trivial insulators with the same $Z_2$ index (0;000) based on parity criteria \cite{Fu_Kane_Z21}.

On the other hand, the opposite parities of the CBM and VBM at the $R$ point suggest that it is possible to drive CsGeBr$_3$ and CsSnBr$_3$ into a topologically nontrivial phase. For CsGeBr$_3$, as shown in Fig.~\ref{fig:CsGeBr$_3$_111_bs_dos} (d), the CBM states mainly come from the Ge $4p$ orbitals, while for CsSnBr$_3$, the CBM states mainly come from the Sn $5p$ orbitals. Therefore, the CBM can be pulled down by substituting the Ge or Sn atoms with foreign atoms that have the similar valence electron configuration as compared to Ge or Sn atoms but lower $p$ orbital level than Ge or Sn atoms. Following this strategy, it is natural to use C atom, which is at the same group as Ge and Sn atoms, as the substitutional dopant. Whereas, the SOC strength of C atom is smaller than those of Ge and Sn atoms, therefore, the C doping will reduce the total SOC strength of the system.

\begin{figure}[tpb]
\includegraphics[width=0.35\textwidth]{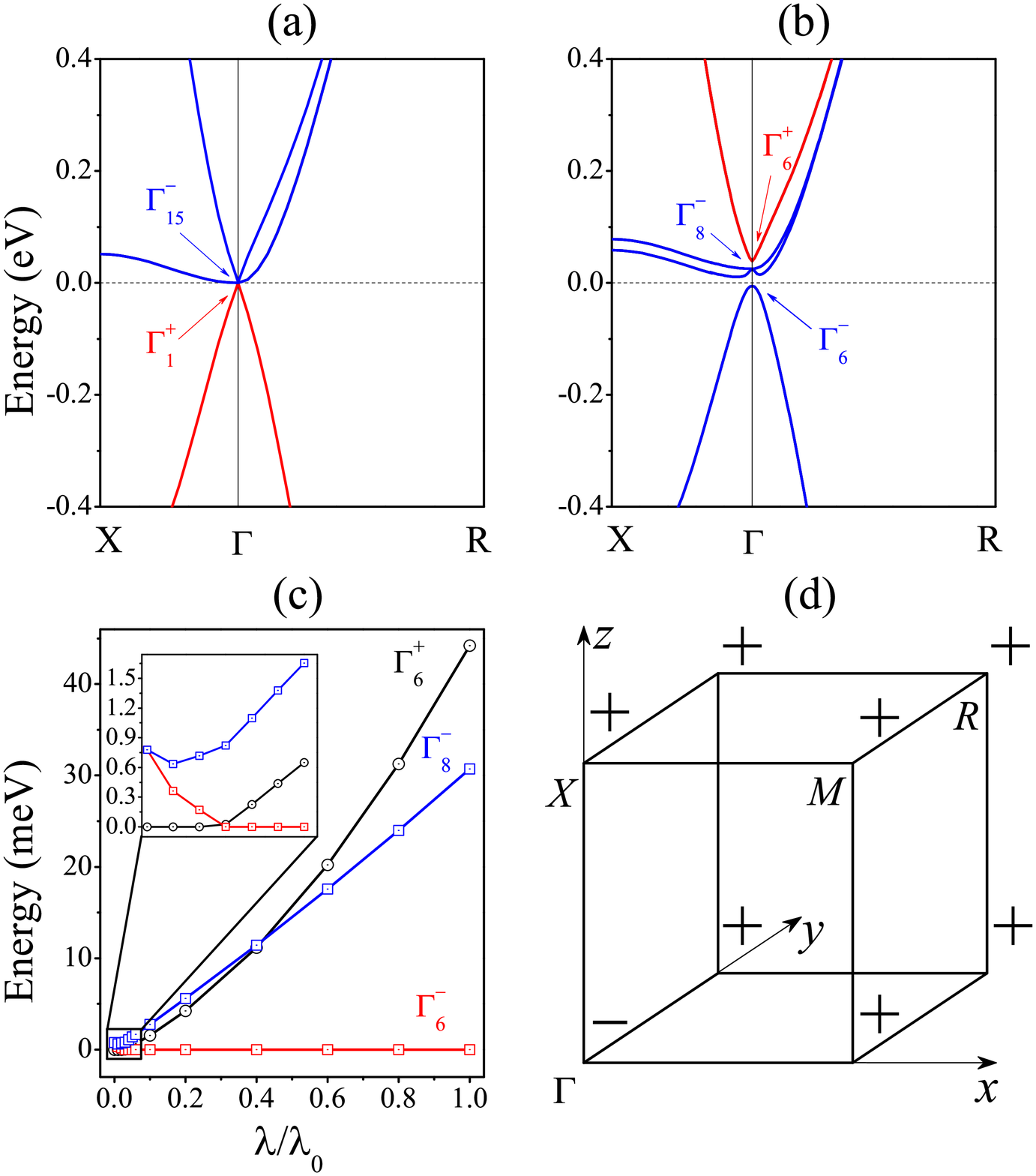}
\caption{\label{fig:c@cgb_bs}(Color online) The PBE calculated electronic band structures of CsGe$_{0.875}$C$_{0.125}$Br$_3$ without (a) and with SOC (b); the calculated energy level of $\Gamma_6^+$, $\Gamma_6^-$ and $\Gamma_8^-$ states (c) as the artificial SOC strength ($\lambda$) increases from zero to original SOC strength ($\lambda_0$); and the diagrams depicting the signs of the products of parity eigenvalues of all the occupied bands at every TRIM point (d). The inset in (c) zooms into the band inversion between $\Gamma_6^+$ and $\Gamma_6^-$. In (a) and (b), the Fermi level is set to zero, and indicated by dashed line.}
\end{figure}

Our PBE calculations show that substitutional C-doping can effectively decrease the band gap and induce a topological phase transition. The results of CsGe$_{0.875}$C$_{0.125}$Br$_3$ are presented in Fig.~\ref{fig:c@cgb_bs}. It is worth noting that the original $R$ point is folded into the $\Gamma$ point because the $2\times2\times2$ supercell is adopted in the calculations. In this case, the VBM and CBM can be re-labeled as $\Gamma_1^+$ and $\Gamma_{15}^-$ states for convenience. After doping C atoms, as expected, the $\Gamma_1^+$ and $\Gamma_{15}^-$ states move toward each other and the band gap decreases close to zero ($\sim$~0.8~meV without SOC) as indicated in Fig.~\ref{fig:c@cgb_bs}~(a). Then we take SOC into account. As shown in Fig.~\ref{fig:c@cgb_bs}~(b), the SOC splits $\Gamma_{15}^-$ states into $\Gamma_8^-$ and $\Gamma_6^-$ states and changes $\Gamma_1^+$ to $\Gamma_6^+$. More importantly, the SOC can further push $\Gamma_6^+$ state up higher than $\Gamma_6^-$ and even $\Gamma_8^-$ states. The detailed evolution of all the states can be further clarified by artificially tuning the SOC strength ($\lambda$) in the calculations from the original SOC strength ($\lambda_0$). As shown in Fig.~\ref{fig:c@cgb_bs}~(c), without SOC, $\Gamma_6^-$ and $\Gamma_8^-$ states are degenerate and higher in energy than the $\Gamma_6^+$ state. By increasing the SOC strength, the splitting between $\Gamma_6^-$ and $\Gamma_8^-$ states increases, and $\Gamma_6^+$ is pushed up relatively. Across a critical point (i.e., $\lambda/\lambda_0 \sim 0.03$), the energy of $\Gamma_6^+$ will be higher than that of $\Gamma_6^-$, and eventually exceed that of $\Gamma_8^-$ when $\lambda/\lambda_0 > 0.4$. The band evolution with SOC implies that there exists a direct-band-gap closing and reopening, which is a solid signal of topological phase transition. To confirm this, we further calculate the products of parity eigenvalues of all the occupied bands at the TRIM points as shown in Fig.~\ref{fig:c@cgb_bs}~(d). According to the parity criteria\cite{Fu_Kane_Z21}, it is clear that the above band inversion only induces a sign change at the $\Gamma$ point and causes the $Z_2$ topological index nontrivial (1;000).

\begin{figure}[htb]
\includegraphics[width=0.35\textwidth]{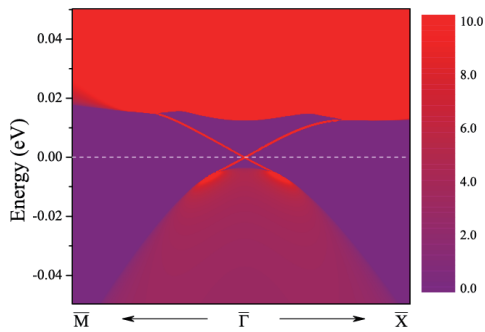}
\caption{\label{fig:c@cgb_ss}(Color online) The calculated LDOS of CsGe$_{0.875}$C$_{0.125}$Br$_3$. The Fermi level is set to zero, and indicated by dashed line. Clearly, Dirac surface states emerge in the bulk gap.}
\end{figure}

The striking properties of 3D topological materials is the Dirac-type surface states in the bulk gap, which can be directly measured by angle resolved photoemission spectroscopy (ARPES). Herein, we calculate the surface states of semi-infinite CsGe$_{0.875}$C$_{0.125}$Br$_3$ by surface Green's function method based on the {\it ab initio} calculation \cite{wannier1, wannier2, GF1}. The imaginary part of the surface Green's function is relative to the local density of states (LDOS), from which we can obtain the surface states. The calculated LDOS of the (100) surface is displayed in Fig.~\ref{fig:c@cgb_ss}. We can clearly see that the topological surface states form a single Dirac cone at the $\Gamma$ point. The Fermi velocity is about 1.45$\times10^5$~m/s, which is of the same order of magnitude as those of Bi$_2$Te$_3$-types TIs\cite{Bi2Se3_exp_3}.

While for CsSnBr$_3$ with Sn substituted by C, the pictures are slightly different. Without SOC, the C doping induces a band inversion: the $\Gamma_{1}^{+}$ states are above the $\Gamma_{15}^{-}$ states and the Fermi level crosses the $\Gamma_{15}^{-}$ states, and therefore, the materials are metallic (see Table~\ref{tab:table}). Although the band inversion is not induced by the SOC, the SOC still plays an important role. Similar to the case of CsGe$_{0.875}$C$_{0.125}$Br$_3$, SOC will split $\Gamma_{15}^-$ into $\Gamma_8^-$ and $\Gamma_6^-$ states and change $\Gamma_1^+$ to be $\Gamma_6^+$, which opens the band gaps around the Fermi level (see Table~\ref{tab:table}) to drive the materials into topologically nontrivial phases. As confirmed by our Z$_2$ calculations, all these materials are 3D strong TIs.

All the above results show that CsGeBr$_3$ with C doping undergoes topological phase transition through path (a)-(b)-(d), while the CsSnBr$_3$ with C doping undergo topological phase transition through path (a)-(c)-(d) (see Fig.~\ref{fig:schem}). Evidently the topological phase transition can be attributed to the lower orbital levels of C $2p$ states in comparison with Ge $4p$ or Sn $5p$ states, and different paths for the topological phase transition originates from the energy difference between the dopant state (i.e., C $2p$ state) and host state (i.e., Ge $4p$ or Sn $5p$ state) around the Fermi level. Larger difference in the orbital levels leads to larger down-shift of the CBM. And the band inversion occurs directly when the energy difference is large enough.

In general, doping smaller atoms will inevitably decrease the lattice constant and have a similar effect as a compressive strain which may induce the topological phase transition in some cases\cite{halide1}. Thus we perform other comparative calculations to provide a full picture of the effects of C doping. When we shrink the lattice constant of CsGeBr$_3$ from the equilibrium lattice of pure CsGeBr$_3$ (5.604~\AA) to that of CsGe$_{0.875}$C$_{0.125}$Br$_3$ (5.543~\AA), the band gap deceases by about 150 meV with or without SOC. On the other side, if we only dope C atoms but fix the lattice constant to the equilibrium lattice of pure CsGeBr$_3$, the band gap decreases by about 630 meV without SOC and 490 meV with SOC, which is still in the topologically trivial phase. The different changes of band gap with and without SOC are due to the smaller SOC strength of C atoms than that of Ge atoms. According to the above comparison, the topological phase transition comes from the combined effects of on-site energy tuning and the lattice shrinkage, while the changes of on-site energy play a dominant role. Similar calculations for CsSn$_{0.875}$C$_{0.125}$Br$_3$ show that even without lattice shrinkage, the band inversion can occur. This again indicates the dominant role of tuning the on-site energy in the topological phase transitions. Note that the substitutional C atoms might shift towards the neighboring Br atoms to increase the C-Br interaction. The resulting structural distortion, though leadings to band splittings due to the break of the lattice symmetry, would not affect the band inversion.

\begin{table}[tpb]
\caption{\label{tab:table2}The HSE calculated band gap at $R$ ($\Gamma$) point without SOC ($E_{\tt g}^{\tt HSE}$) and with SOC ($E_{\tt g}^{\tt HSE-soc}$) of pure CsGeBr$_3$ and CsSnBr$_3$ (CsGe$_{0.875}$C$_{0.125}$Br$_3$ and CsSn$_{0.875}$C$_{0.125}$Br$_3$). ``$-$''~presents the inverted band gap. }
\begin{ruledtabular}
\begin{tabular}{ccc}
                                & $E_{\tt g}^{\tt HSE}$ (meV) & $E_{\tt g}^{\tt HSE-soc}$ (meV) \\
\colrule
CsGeBr$_3$                      & 1246.5 & 1068.2 \\
CsGe$_{0.875}$C$_{0.125}$Br$_3$ & 363.3  & 318.1  \\
CsGe$_{0.875}$C$_{0.125}$Br$_3$\footnote{with 7\% volume decreasing} & 0.0  & $-$64.5  \\
CsSnBr$_3$                      & 1302.8 & 927.8  \\
CsSn$_{0.875}$C$_{0.125}$Br$_3$ & 17.6   & $-$62.2\\
\end{tabular}
\end{ruledtabular}
\end{table}

It is well known that PBE usually underestimates the band gap due to the self-interaction error. Specifically, the band gap of cubic CsGeBr$_3$ is 1.59 eV (at 300~K under 1.2~GPa pressure) as measured by optical absorption\cite{CGB-temperature-pressure}. The band gap of CsSnBr$_3$ is 1.80~eV (0.34~eV) for the cubic phase at 300~K as obtained from optical (transport) experiments\cite{CSB-temperature-BG}. The PBE calculated band gap of the cubic-phase CsGeBr$_3$ and CsSnBr$_3$ are 583.4 and 279.2~meV (including the spin-orbit coupling), respectively (see Table \ref{tab:table}). We employ HSE functional to check the band gap issue. For the CsGeBr$_3$ and CsSnBr$_3$ with SOC, the HSE predicted band gaps are 1068.2~meV and 927.8~meV, respectively. Although they are still smaller than the experimental values and $GW$ calculated results\cite{CSB-GW}, they are much improved over PBE calculated ones. For CsGe$_{0.875}$C$_{0.125}$Br$_3$ and CsSn$_{0.875}$C$_{0.125}$Br$_3$, the HSE calculated band gaps are 318.1~meV and $-62.2$~meV (the negative sign denotes a band inversion), respectively. This suggests that CsSn$_{0.875}$C$_{0.125}$Br$_3$ is a TI, while CsGe$_{0.875}$C$_{0.125}$Br$_3$ is not. Despite that CsGe$_{0.875}$C$_{0.125}$Br$_3$ is topologically trivial, the substitution of Ge with C shifts the CBM downward significantly. Applying an external pressure could help drive CsGe$_{0.875}$C$_{0.125}$Br$_3$ into TI. The HSE calculations predict that the band order would get inverted if decreasing the material volume by 7\%, and a $-64.5$~meV band gap is obtained when including the SOC (see Table~\ref{tab:table2}).
From the above PBE and HSE calculations, one knows that even using the advanced methods, there could exist a discrepancy in the band gap between the prediction and the reality, and an accurate description of band gap is quite challenging and beyond the scope of the present work. However, importantly, by using different exchange-correlation functionals we get the same physical picture that the substitution of Ge or Sn with C tunes orbital levels and thus facilitates a band inversion, validating the concept we proposed. Note that the band gap at the $\Gamma$ point predicted by HSE is $\sim$20~meV larger than that by PBE (see Tables I and II). This is presumably because the (partial) correction of the self-interaction error in HSE, which may lead to relatively more localized electronic states than PBE, could give a stronger effective SOC.

It should be noticed that the temperature effects are important to the present system. As the band gap of the TI phase is on the order of the room-temperature thermal excitation energy of 26~meV, the thermal excitation would affect transport properties of metallic surface states by some extent. Nevertheless, the existence of massless Dirac fermions on the surface, as a hallmark of topological insulators, is expected to be detectable by ARPES. Another intriguing feature must be mentioned, lowering the temperature could result in a TI to normal insulator transition, as caused by a temperature-induced structural phase transition which can increase the band gap. This unusual feature will be discussed later in detail in our future work.

Based on the general tight-binding picture, it is expected that the heavy elements doping can also induce the topological phase transition by pulling up the VBM and increasing the SOC strength in some materials. Furthermore, our strategy of tuning the onsite energy is not limited for 3D TIs. As well known, the band inversion also has a similar fundamental effect on weak TIs, topological crystalline insulators \cite{TCI1, TCI2, TCI3}, the quantized anomalous Hall insulators \cite{QAH1, QAH2} and other topological systems. For all those systems, in principles, we can tune the on-site energy by doping and make them nontrivial.

In summary, we propose a strategy of realizing topological phase transition by tuning orbital levels via chemical doping. Due to the different orbital level of dopant and the substituted host atoms, the dopant can affect the energy band around the Fermi level and can reduce the band gap effectively or even induce a band inversion, and promote the topological phase transition.

This work was supported by the Ministry of Science and Technology of China (Grant Nos. 2011CB606405 and 2011CB921901) and the National Natural Science Foundation of China (Grant No. 11334006). The calculations were performed on the IBM Blade cluster system in the High Performance Computing Center at Nanjing University and the ``Explorer 100'' cluster system at Tsinghua University. J. L. acknowledged support from the STC Center for Integrated Quantum Materials, NSF Grant No. DMR-1231319.

\end{document}